\documentclass[prd,aps,twocolumn,nofootinbib,preprintnumbers,superscriptaddress,preprintnumbers,balancelastpage,longbibliography]{revtex4-2}

\usepackage{graphicx}
\usepackage{float}
\usepackage{bm}
\usepackage{times}
\usepackage{slashed}
\usepackage{color}
\usepackage{aas_macros}
\usepackage{slashed}
\usepackage{lipsum}
\usepackage{subfigure}
\usepackage{multirow}
\usepackage{amsmath}
\usepackage{array} 
\usepackage{varwidth} 
\usepackage{hyperref}
\usepackage[dvipsnames]{xcolor}
\usepackage{listings}
\usepackage{color,xcolor}

\newcommand{\be}{\begin{equation}}
\newcommand{\ee}{\end{equation}}
\newcommand{\bea}{\begin{eqnarray}}
\newcommand{\eea}{\end{eqnarray}}

\hypersetup{
     colorlinks   = true,
     citecolor    = blue,
     urlcolor     = blue,
     linkcolor    = blue
}

\begin{document}

\title{Strong Constraints on Line Signals from Dark Matter Annihilation in a Nearby Subhalo}

\author{Asier Salces Pérez}
\email{asiersperez@gmail.com, ORCID: orcid.org/0009-0006-9113-1867}
\affiliation{Stockholm University and The Oskar Klein Centre for Cosmoparticle Physics,  Alba Nova, 10691 Stockholm, Sweden}
\affiliation{Departamento de F\'{\i}sica Te\'orica, M-15, Universidad Aut\'onoma de Madrid, E-28049 Madrid, Spain}
\affiliation{Instituto de F\'{\i}sica Te\'orica UAM-CSIC, Universidad Aut\'onoma de Madrid, C/ Nicol\'as Cabrera, 13-15, 28049 Madrid, Spain}

\author{Thong T.Q. Nguyen}
\email{thong.nguyen@fysik.su.se, ORCID: orcid.org/0000-0002-4436-0820}
\affiliation{Stockholm University and The Oskar Klein Centre for Cosmoparticle Physics,  Alba Nova, 10691 Stockholm, Sweden}

\author{Pedro De la Torre Luque}
\email{pedro.delatorre@uam.es, ORCID: orcid.org/0000-0002-4150-2539}
\affiliation{Departamento de F\'{\i}sica Te\'orica, M-15, Universidad Aut\'onoma de Madrid, E-28049 Madrid, Spain}
\affiliation{Instituto de F\'{\i}sica Te\'orica UAM-CSIC, Universidad Aut\'onoma de Madrid, C/ Nicol\'as Cabrera, 13-15, 28049 Madrid, Spain}

\author{Tim Linden}
\email{linden@fysik.su.se, ORCID: orcid.org/0000-0001-9888-0971}
\affiliation{Stockholm University and The Oskar Klein Centre for Cosmoparticle Physics,  Alba Nova, 10691 Stockholm, Sweden}

\begin{abstract}
\noindent We present a dedicated search for monochromatic gamma-ray emission from a recently proposed nearby dark matter subhalo candidate. Using nearly 15 years of Fermi-LAT Pass 8 data, we perform a sliding-window search for gamma-ray lines between 10 and 300 GeV. We do not find any statistically significant evidence for a line signal. The largest excess occurs at $E_\gamma \simeq 28 \, {\rm GeV}$ with a local significance of $2.5\sigma$. We therefore derive 95\% confidence level upper limits on the annihilation cross section for $\chi\chi \rightarrow \gamma\gamma$. Under the assumed NFW subhalo model, the resulting constraints are stronger than existing Galactic Center line limits over much of the explored mass range. We further translate these line constraints into bounds on well-motivated WIMP models, such as Higgs-portal and Wino dark matter scenarios, restricting previously open parameter regions consistent with thermal freeze-out as well as models associated with dark matter interpretations of the Galactic Center Excess.
\end{abstract}

\maketitle

\section{Introduction}

Dark matter (DM) constitutes most of the matter density in the universe, producing potential wells that drive cosmic structure formation. Its existence is supported by a wide range of observations at all scales \cite{Cirelli:2024DarkMatter, Bertone:2004pz, Clowe:2006eq}, from galaxy rotation curves \cite{rubin1970rotation} to the cosmic microwave background~\cite{Planck:2018vyg}. Despite its gravitational interaction, the particle nature of DM remains unknown. Among the most well-motivated candidates are weakly interacting massive particles (WIMPs). Their thermal production in the early universe can reproduce the measured abundance with annihilation cross sections consistent with weak forces \cite{Bertone:2004pz, Jungman:1995df}. Since the same annihilation process still occurs at late times in regions of high DM density, indirect searches for the stable Standard Model particles produced in these annihilations can strongly probe WIMP parameters~\cite{Bergstrom:2000pn}.

WIMP DM could annihilate into many different Standard Model final states, producing observable cosmic-ray, neutrino and gamma-ray signals. Many of these tree-level annihilation channels create broad gamma-ray spectra coming from the decay or hadronization of final-state particles, such as \mbox{$\chi\chi \to b\bar{b}$}, \mbox{$\chi\chi \to \tau^+\tau^-$}, etc \cite{Cirelli:2010xx}. However, DM may also annihilate directly into two-body final states containing photons, like $\gamma\gamma$ or $\gamma Z$, producing monochromatic spectral lines fixed by the DM mass \cite{Bergstrom:1988fp, Rudaz:1989ij}. Although these monochromatic lines usually arise only at loop level and are suppressed relative to tree-level annihilation channels, their distinct spectra make them one of the cleanest signatures for indirect detection. In specific scenarios, these signals can be enhanced, for example through resonant annihilation \cite{Hisano:2004ds} or internal bremsstrahlung from nearly degenerate charged states \cite{Bringmann_2008}.

\begin{figure}[t!]
    \centering
    \includegraphics[width=1\linewidth]{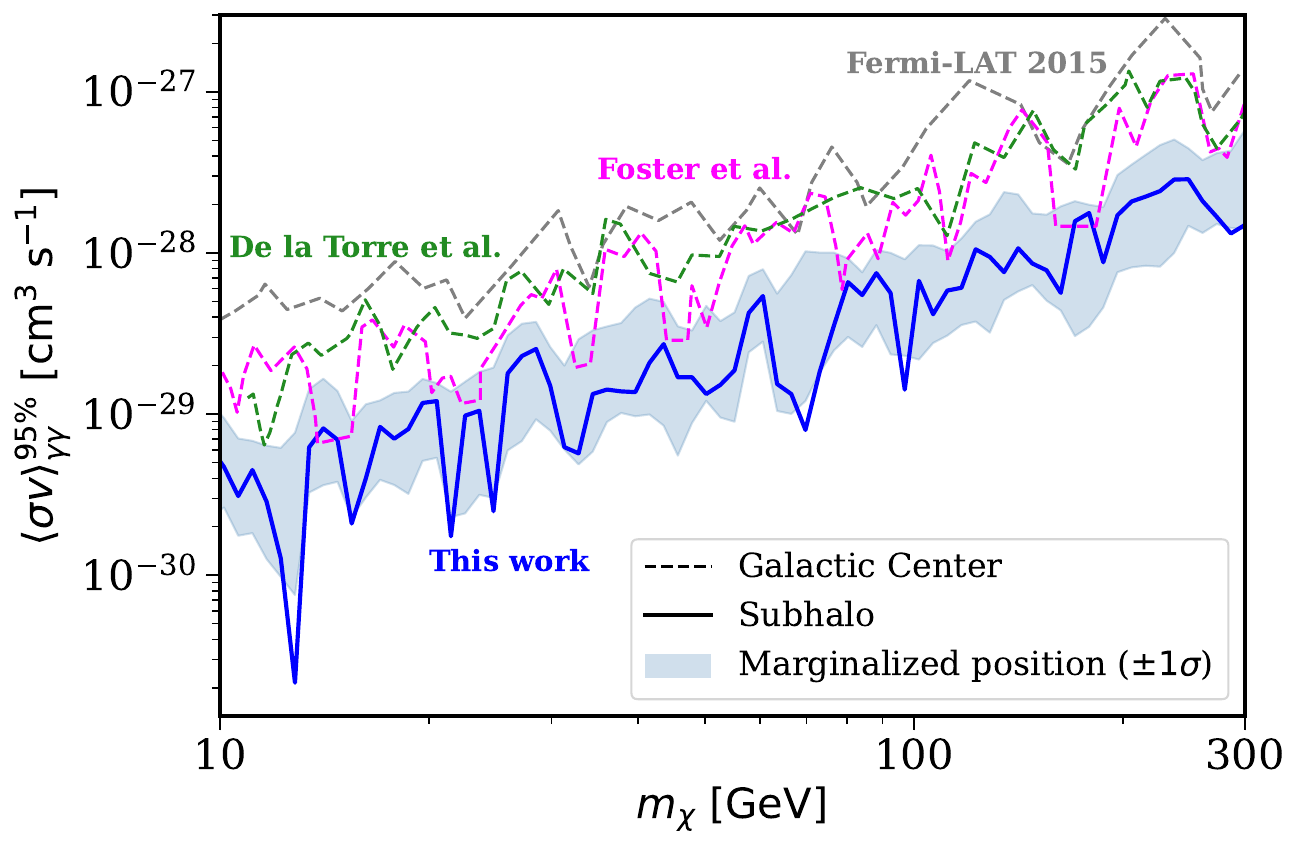}
    \caption{Constraints on DM annihilation cross section for $\chi\chi\to \gamma\gamma$ annihilation, assuming an NFW profile. Our 95\% CL limits using Fermi-LAT subhalo observation are in blue. Previous results for single-line searches using Galactic Center observation for comparison are: Fermi-LAT (2015) is in gray~\cite{Fermi-LAT:2015kyq}, De la Torre Luque et al. (2023) is in green~\cite{luque2023gammaraylines15years}, and Foster et al. (2022) is in magenta~\cite{Foster_2023}.}
    \label{fig:boundscomparison}
\end{figure}

The Galactic Center (GC) has long been considered as one of the most promising targets for gamma-ray line searches with Fermi-LAT data, due to the large DM annihilation signal produced by the high DM density expected in the inner Milky Way \cite{Navarro:1995iw}. Beyond the GC, similar strategies have also targeted other DM rich systems, including dwarf spheroidal galaxies, galaxy clusters and nearby external galaxies, which provide complementary analysis in regions with large DM ~\cite{McDaniel:2023bju}.  Early studies reported a tentative line-like feature near $133\,{\rm GeV}$ with a global significance of $3.2\sigma$ \cite{weniger2012tentative}. A contemporaneous stacking analysis of eighteen nearby galaxy clusters also reported a double feature near 110 and 130 GeV, with a global significance of up to $3.6\sigma$ \cite{Hektor:2012kc}. However, the official Fermi-LAT reanalysis, improving the treatment of energy dispersion, found the feature near $133\,{\rm GeV}$ to be reduced to a global significance of $1.5\sigma$. It was also seen that this excess had to be narrower than the expected LAT resolution, disfavoring its interpretation as a physical gamma-ray line \cite{Fermi-LAT:2013thd}. A subsequent analysis using Pass 8 data did not find significant line detection and did not confirm the previous features \cite{Fermi-LAT:2015kyq}. Later Pass 8 searches towards galaxy clusters found no globally significant line signal, although a tentative line-like excess near $43\,{\rm GeV}$ was reported in one analysis \cite{Anderson:2015dpc, Liang:2016pvm}.

More recent works using 14 years of Fermi-LAT observations, together with improved spectral and instrumental treatments, have not found significant gamma-ray lines and have placed strong constraints on $\langle \sigma v \rangle_{\gamma\gamma}$ \cite{Foster_2023, luque2023gammaraylines15years}. At higher energies, searches with H.E.S.S. and MAGIC have extended the mass range up to $100\,{\rm TeV}$, without detecting any significant excess \cite{HESS:2018cbt, MAGIC:2022acl}. A recent stacking analysis of 15.5 years of Fermi-LAT data has reported a renewed signal near $43.2\,{\rm GeV}$ in the directions of the Virgo, Fornax and Ophiuchus clusters \cite{Fan:2024rcr}. However, the absence of this feature in the inner Galaxy disfavors a DM annihilation interpretation. Therefore, no gamma-ray line has yet been interpreted as a DM signal. The field is currently characterized by strong upper limits in the GeV-TeV range with some target-dependent anomalies.

Even though the GC is expected to provide an extremely high dark matter flux, it is a complex astrophysical environment, and the interpretation of GC limits remains affected by significant uncertainties in the DM density profile of the inner Galaxy \cite{Abdo:2010nc}. In gamma-ray line searches, the impact of the background is minimized, but the conversion of a line-flux limit into a constraint on the annihilation cross section remains strongly dependent on the assumed $J$-factor, and hence on the inner DM density profile. In contrast, nearby DM-dominated systems provide complementary targets for indirect searches. Dwarf spheroidal galaxies are the standard example, because they have low astrophysical backgrounds and their \(J\)-factors can be estimated from stellar kinematics~\cite{Strigari_2008, crnogorcevic2024strongconstraintsdarkmatter, Baltz_1999,Geringer_Sameth_2015}. Nearby dark subhalos could offer a similar advantage. In the cold dark matter scenario, N-body simulations predict that Galactic halos contain a large population of bound substructures, some of which may remain dark because they have not formed enough stars to be detected as luminous satellites~\cite{Springel:2008cc, Diemand:2008in}. Such objects could potentially provide large annihilation fluxes in cleaner gamma-ray environments than the GC. The main challenge, however, is that these sources are usually identified through blind searches in gamma-ray catalogs: without independent dynamical information, their distance, mass, and \(J\)-factor remain degenerate and must be inferred statistically from subhalo population models~\cite{di2020investigating, Crnogorcevic:2023ijs}.

Recently, Chakrabarti et al.~\cite{29xz-nt5z} addressed this limitation by introducing a method based on pulsar timing acceleration measurements, using them as probes of local deviations from a smooth Galactic gravitational potential. They applied this strategy and discovered a localized gravitational anomaly in the Milky Way, obtaining independent dynamical information on the location and mass of the object, rather than relying only on its gamma-ray properties. This gravitationally selected candidate therefore represents a rare and potentially unique target for indirect DM searches. Their study favors an interpretation where the gravitational anomaly is due to DM substructure, obtaining Bayes factors of order $\sim20-40$ suggesting tentative evidence for the subhalo scenario. Using an NFW profile, they fix the scale radius and concentration to $r_s = 0.1\,{\rm kpc}$ and $c = 30,$ respectively, adopting a compact profile motivated by cosmological expectations for low-mass CDM subhalos \cite{Bullock_2017}. With these magnitudes, they infer a virial mass of:
\begin{equation}
     M_{\rm sub} = \left(6.19^{+1.92}_{-2.03}\right)\times 10^{7}\,M_{\odot}
\end{equation}
and place the center of the object at the Galactocentric coordinates:
\begin{equation}
X = 7.47^{+0.21}_{-0.14}\,{\rm kpc}, 
\;
Y = 0.38^{+0.11}_{-0.16}\,{\rm kpc}, 
\;
Z = 0.21^{+0.06}_{-0.11}\,{\rm kpc}
\end{equation}
yielding a heliocentric distance of $d = 0.783^{+0.129}_{-0.194} \text{ kpc}$ using $R_\odot = 8.122\,{\rm kpc}$ \cite{GRAVITY:2018ofz}. 
A subsequent analysis by Zhu et al.~\cite{zhu2025constraintsdarkmatterannihilation} searched for broadband gamma-ray emission from the same gravitationally selected subhalo candidate using Fermi-LAT data. They modeled the expected emission as an extended NFW template and propagated the geometric uncertainty by considering the fiducial heliocentric distance together with its $1\sigma$ lower and upper values, which modifies both the angular morphology and the J-factor normalization. They reported a tentative residual excess near the candidate direction. Since its origin was unclear, they kept it in the analysis rather than masking it or fitting it with an additional source. This weakens the resulting limits and therefore makes them conservative, because the dark matter component is allowed to account for part of the excess when deriving limits on continuum annihilation channels \(b\bar b\) and \(\tau^+\tau^-\). These results motivate this work as a complementary search for monochromatic gamma-ray lines on the same target.

In this paper, we perform a dedicated search for monochromatic gamma-ray lines from the proposed nearby subhalo candidate. Motivated by previous line searches in the Galactic Center and dwarf
spheroidal galaxies, as well as by the recent continuum analysis of this source, this work provides an additional test focused on narrow spectral features. In this regime, where the exact spatial morphology and position of the source remain uncertain, a line search is especially useful because it relies primarily on a spectral peak rather than on a detailed spatial template. Our analysis uses a sliding-window approach applied to Fermi-LAT data in the region of interest around the source. We find a local excess with a local significance of \(2.5\sigma\) best-fit by a dark matter candidate with a mass of approximately \(\sim 28\,{\rm GeV}\), which is not sufficient to claim a detection but motivates a careful characterization of the candidate line feature. We also derive upper limits on the annihilation cross section into \(\gamma\gamma\) and construct expected sensitivity bands from background-only pseudo-experiments.

This paper is organized as follows: In Sec.~\ref{sec:methodology}, we describe the Fermi-LAT dataset used in this work and the sliding-window strategy adopted for the gamma-ray line search. In Sec.~\ref{sec:Constraints}, we compute the $J$-factor of the subhalo candidate and show both the Test Statistic and the constraints on the annihilation cross section, including implications for some well-motivated WIMP DM scenarios. Finally, Sec.~\ref{sec:conclusions} presents the main conclusions of this work.

\section{METHODOLOGY AND DATA ANALYSIS}
\label{sec:methodology}
\subsection{Data treatment}
For this work, we use the nearly $\sim$15 years of Fermi-LAT PASS8 data (2008-08-04 to 2023-07-20) that were extracted in Ref.~\cite{luque2023gammaraylines15years}. We select P8R3\_CLEAN events (\texttt{evclass}=256) in the energy range between \mbox{300~MeV--300~GeV.} We apply a zenith angle cut to remove events with zenith angles $z$$>90^{\circ}$ in order to avoid contamination coming from Earth's limb. For the line analysis, we select events within a circular ROI of $10^\circ$ radius centered on the subhalo position, as motivated below.
In this analysis, we keep all PASS8 event types without separately dividing them into ``EDISP" classes, given that the impact in the final result is expected to be mild. We account for the energy dispersion of the selected CLEAN event sample through the corresponding Fermi-LAT energy dispersion response matrix. This matrix encodes the probability of reconstructing a photon with true energy $E_{\rm true}$ at a measured energy $E$, and is derived from the PASS8 instrumental response functions.
The extraction of the Fermi-LAT data exposure maps and energy dispersion matrix is done with the official ScienceTools \cite{Fermitools}, and we do not apply any masks to remove astrophysical emission from either diffuse or point sources.

We set the subhalo at a heliocentric distance of \mbox{$d = 0.783^{+0.129}_{-0.194}~\text{ kpc}$} and assume a subhalo scale radius of $r_s = 0.1\, {\rm kpc}$, which produces a scale angle $\theta_s = \arctan(r_s/d)$ that varies from $\sim6^\circ$ to $\sim10^\circ$. Therefore, it is justified to adopt a circular ROI of $10^\circ$ radius centered in the NFW best-fit coordinates of the subhalo. To propagate the uncertainty in the subhalo position, we scan 300 realizations by independently moving each Galactocentric coordinate, $x$, $y$ and $z$ within $\pm 2\sigma$ of its fiducial value. We assign a Gaussian weight to each realization according to its position
\begin{equation}
    w_i =
\frac{
\exp\left[-\frac{1}{2}\left(
\frac{(X_i-X_0)^2}{\sigma_X^2}
+\frac{(Y_i-Y_0)^2}{\sigma_Y^2}
+\frac{(Z_i-Z_0)^2}{\sigma_Z^2}
\right)\right]
}{
\sum_j
\exp\left[-\frac{1}{2}\left(
\frac{(X_j-X_0)^2}{\sigma_X^2}
+\frac{(Y_j-Y_0)^2}{\sigma_Y^2}
+\frac{(Z_j-Z_0)^2}{\sigma_Z^2}
\right)\right]
}.\label{eq:gaussweight}
\end{equation}
and the corresponding heliocentric distance and $J$-factor are recomputed.

\subsection{Monochromatic line analysis}
We follow the same strategy as in Fermi-LAT collaboration searches for spectral lines \cite{Abdo:2010nc} by producing a maximum likelihood fit in the $20^\circ\times20^\circ$ ROI, scanning over 88 sliding energy intervals (using the method of Ref.~\cite{GillesVertongen_2011}) from 10 GeV to 300 GeV. We include all events that fall within the ROI without any additional spatial weighting. For each trial energy $E_\gamma$, we fit the photon counts within an energy window of half-width $3\sigma_E(E_\gamma)$, where $\sigma_E$ is the 68\% exposure-weighted energy resolution for the dataset used in this analysis (see Ref.~\cite{LATPerformance}). For each energy window, we model the background with a power-law $n_\mathrm{bkg}(E) = A \left(E/E_\gamma\right)^{-\gamma}$ allowing both the normalization and the spectral index to vary freely. We build a line-like signal as a generic monochromatic excess at energy $E_\gamma$, with $n_s$ as the only signal parameter, representing the total number of line photons in the ROI. The likelihood function we use is described as a Poisson distribution for the number of counts in all the energy windows
\begin{equation}
    \mathcal{L}_\mathrm{ROI} = \prod_i \frac{e^{-n(E_i, E_i')} \cdot n(E_i, E'_i)^{N_i(E_i)}}{N_i!},
    \label{likelihood}
\end{equation}
where $N_i$ is the observed number of photons in energy bin $E_i$ within the ROI, and $n(E_i, E_\gamma)$ is the expected number of counts at true energy $E_\gamma$ reconstructed at energy $E_i$ by the instrument. We first fit a null hypothesis, where the number of signal events is 0 and the expected $\gamma$-ray counts spectrum is obtained by fitting the data to a power-law. The alternative hypothesis, on the other hand, describes the number of counts as a sum of the background plus the non-negative line signal convolved with the energy dispersion matrix $n(E_i, E_\gamma) = n_{bkg} + n_s  \mathcal{E}(E_\gamma, E_i)$. The form of the latter term comes from the fact that the LAT measures photon energies with finite resolution, meaning that a monochromatic line at $E_\gamma$ is smeared across close energy bins \cite{FermiPass8Edisp}. The matrix $\mathcal{E}(E_\gamma, E_i)$ accounts for this by encoding the probability of a photon emitted at $E_\gamma$ being reconstructed at $E_i$. This prescription follows previous studies and recommendations from the Fermi-LAT collaboration~\cite{Fermi-LAT:2015kyq, Abdo:2010nc}.

The spectral fits are performed using Markov Chain Monte Carlo (MCMC) with the \textit{emcee} ensemble sampler \cite{Foreman_Mackey_2013}. This approach is more robust because it allows us to explore the posterior distribution of both the background and signal parameters, rather than relying only on a best-fit solution from conventional optimizers. These probability distributions are then used to derive credible intervals for the fitted quantities and to obtain the corresponding DM limits.

The best-fit number of counts can be small, especially in energy windows affected by downward background fluctuations. This can lead to a mis-evaluation of the confidence intervals, in particular due to the fact that the standard confidence interval may include non-physical negative values of the line-strength. To correctly calculate the strength of the upper limit in scenarios where the best-fit number of counts is negative, we utilize the Feldman-Cousins (FC) construction \cite{PhysRevD.57.3873} \footnote{The package is obtained from \href{https://github.com/usnistgov/FCpy/tree/main}{https://github.com/usnistgov/FCpy/tree/main}}. In practice, we take the best-fit from the MCMC posterior and apply the FC method with the likelihood function with Eq.~\ref{likelihood} where the best-fit number of signal counts is below $n_s = 2$.

\section{DM CONSTRAINTS FROM THE LINES}
\label{sec:Constraints}
In order to obtain the constraints, we compute the expected gamma-ray flux coming from a $\chi\chi \to \gamma\gamma$ process within the subhalo:
\begin{equation}
    \frac{{\rm d}\Phi}{{\rm d}E}
=
\frac{1}{8\pi}
\frac{\langle\sigma v\rangle_{\gamma\gamma}}{m_{\chi}^{2}}
\left(\frac{dN}{dE}\right)_{\gamma\gamma}
\mathcal{J}_{\rm ROI}(\Delta\Omega),
\label{limits}
\end{equation}
where $m_\chi$ is the mass of the DM particle and \mbox{$\frac{dN}{dE}_{\gamma\gamma} = 2\delta(E-E_{\gamma\gamma})$} is the spectrum of gamma-rays per interaction, where the factor of 2 accounts for the fact that two photons are produced in each annihilation event, and $\mathcal{J}_{ROI}(\Delta\Omega)$ is the astrophysical J-factor of the subhalo:
\begin{equation}
    \mathcal{J}_{ROI} = \int_0^{\theta_{\rm max}} 2\pi \sin\theta \, d\theta 
\int_{-l_{\rm max}}^{l_{\rm max}} \rho^2\!\left(r(l,\theta)\right) dl,
\end{equation}
where the 3D distance from the subhalo center is
\begin{equation}
    r(l, \theta) = \sqrt{l^2 + d_\odot^2 - 2\, l\, d_\odot \cos\theta},
\end{equation}
where $l$ denotes the coordinate along the line of sight, $d_\odot$ is the heliocentric distance to the subhalo and $\theta$ is the angular separation from the subhalo direction. For the fiducial model, in which the source is left unchanged, $d_\odot = 0.783$ kpc. For shifted positions, their corresponding heliocentric distance is calculated. We model the subhalo density with an NFW profile in order to be consistent with the analysis adopted by Chakrabati et al.~\cite{29xz-nt5z}. In the fiducial set-up, we obtain a $\mathcal{J}_{ROI} = 1.151\times10^{23} \, {\rm GeV^2/cm^5}$

\subsection{Looking for lines in the gamma-ray spectrum}
In order to calculate the local significance of the lines, we compute the Test Statistic (TS) from the likelihood ratio,
\begin{equation}
\mathrm{TS} = -2\ln\left[\frac{\mathcal{L}(A, \gamma)}{\mathcal{L}(A, \gamma, n_s)}\right] = -2\left[\ln\mathcal{L}_\mathrm{NoS} - \ln\mathcal{L}_\mathrm{S}\right],
\end{equation}
where the numerator is the best-fit likelihood under the null hypothesis (background only) and the denominator is the best-fit (maximal) value under the alternative hypothesis (background + DM signal). Assuming that Wilks' theorem holds, the local significance is then estimated as $\sigma_{local} = \sqrt{TS}$ for positive best-fit signal normalizations, and is set to zero otherwise. It is important to remark that, although the dominant uncertainty in the expected annihilation signal is the subhalo position and hence its associated J-factor, the TS itself is independent of this magnitude. since it only tests the presence of a line-like excess in the data independently of the assumed DM normalization.

\begin{figure}[!t]
    \centering    \includegraphics[width=1\linewidth]{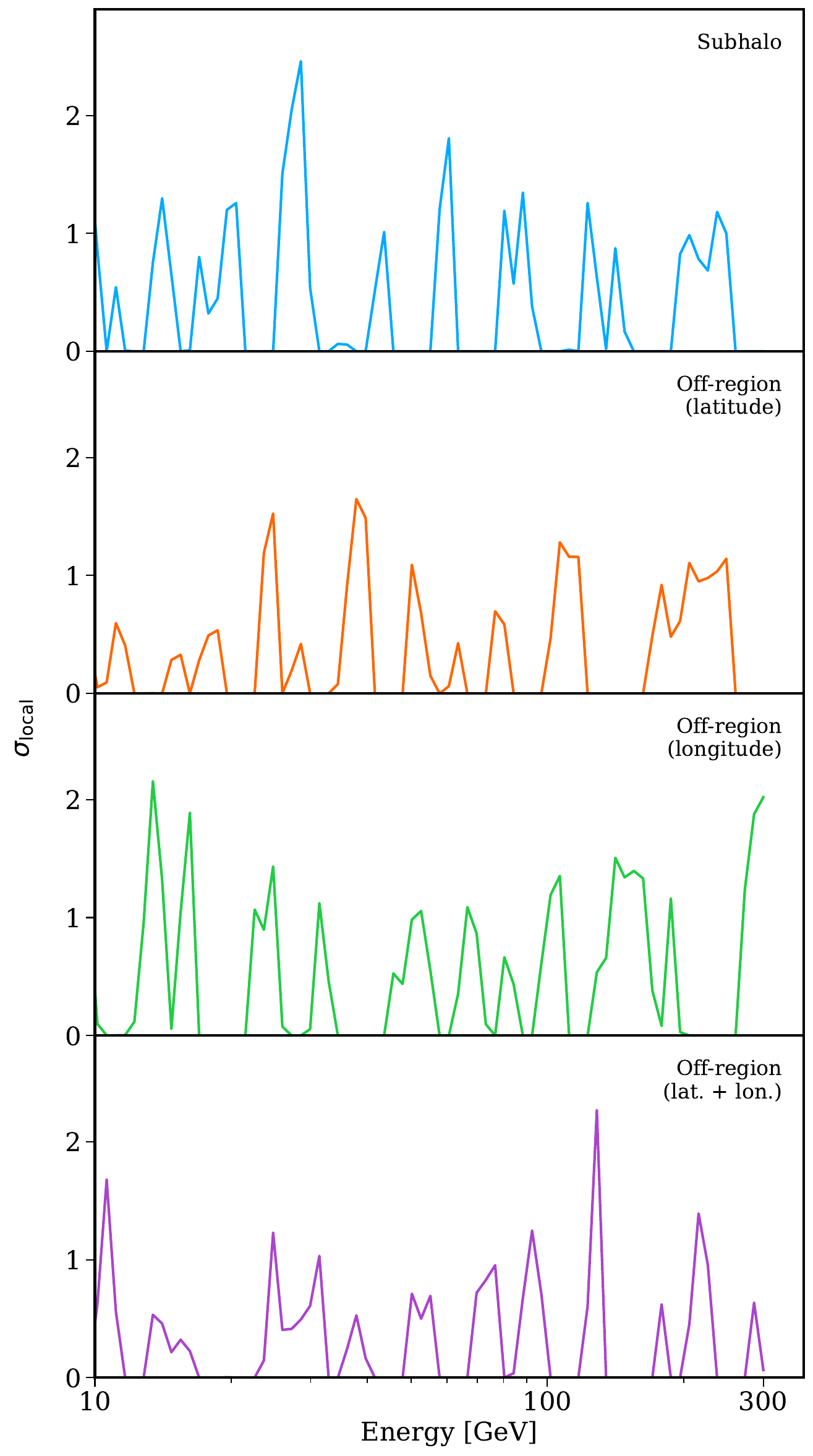}
    \caption{The local significance obtained in the 10--300 GeV range. The result of the best fit subhalo position is compared with three off regions. These locations are reflections in latitude, longitude and in both latitude and longitude, where we do not expect flux coming from the subhalo. The $\sigma_{local}$ is set to 0 when the number of counts that maximizes the likelihood is not a positive value.}
    \label{fig:sigmaoff}
\end{figure}

Fig.~\ref{fig:sigmaoff} shows the results of the line search for the fiducial subhalo position together with the corresponding off-regions that are reflections in Galactic latitude and longitude. In the subhalo direction, the largest feature appears at approximately $E_\gamma\simeq28$ GeV, reaching a local significance of about $2.5\sigma$. We do not notice any comparable excess at the same energy, supporting the interpretation that the feature is localized around the subhalo direction rather than being a generic artifact of the smooth background model or instrumental response. Nevertheless, given its modest local significance, we regard this feature as an interesting excess, but not as evidence for a gamma-ray line detection.

As a further robustness check, we repeat the line search for several subhalo positions shifted by $\pm1\sigma$ in the Galactocentric coordinates to see if this peak vanishes with the position. The results are shown in Appendix~\ref{sec:additionalp}.

\subsection{DM constraints on the monochromatic line analysis}
Given that we do not find statistically significant gamma-ray line, we use the results to derive 95\% confidence level upper limits on the annihilation cross section $\langle\sigma v\rangle_{\gamma\gamma}$. For each tested line energy, the 95th percentile of the posterior $n_s$ distribution is converted into a flux limit using the exposure at the corresponding energy and then into a cross-section limit using Eq.(\ref{limits}).

\begin{figure}[!t]
    \centering
    \includegraphics[width=1\linewidth]{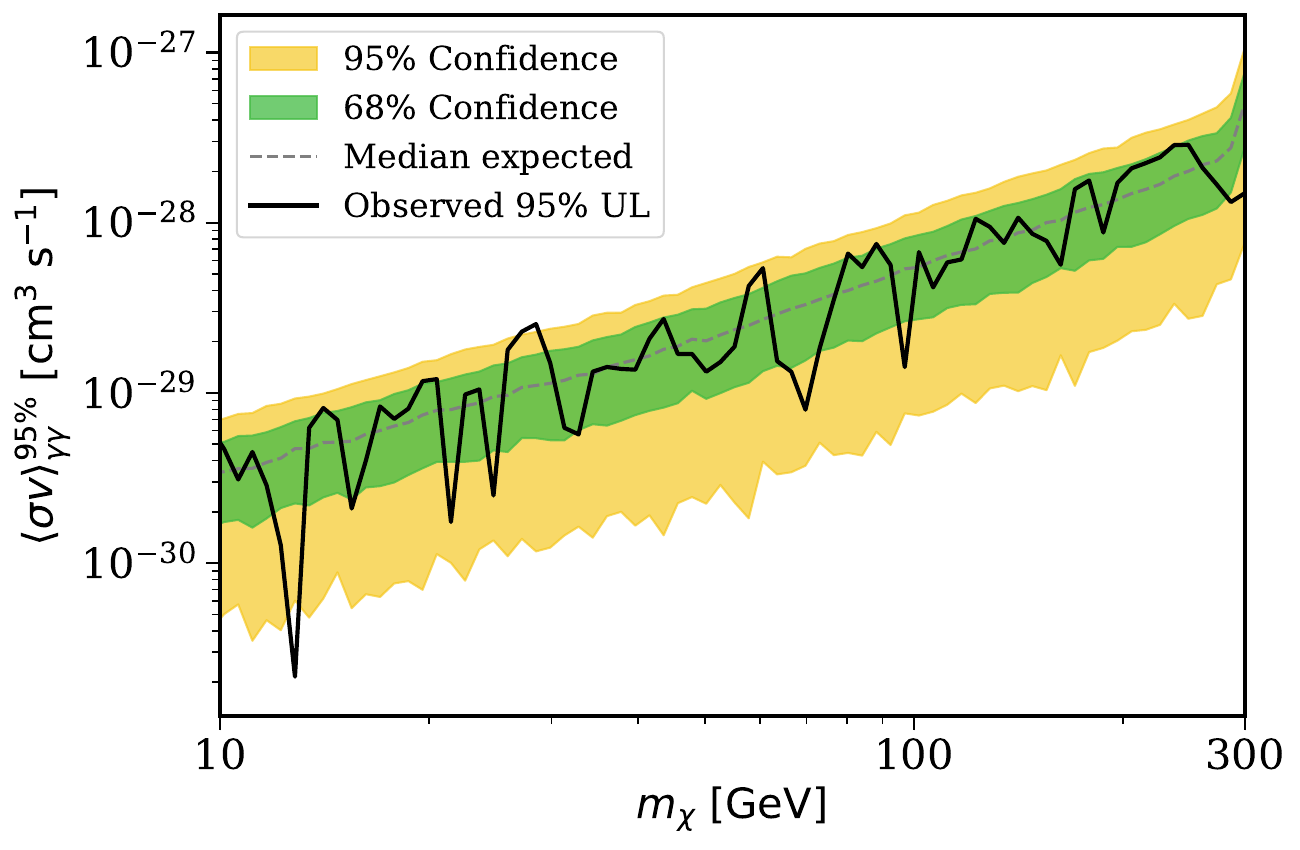}
    \caption{95\% confidence level upper limits on the DM annihilation cross section $\langle\sigma v\rangle_{\gamma\gamma}$ for the subhalo assuming a NFW profile. The solid line shows the results for the line searches while the green and yellow shaded regions show the 68\% and 95\% containment bands respectively.}
    \label{fig:limitswithbands}
\end{figure}

To estimate the expected sensitivity of the analysis, we generate mock data sets from a smooth power-law model that best fits the observations in the region, including Poisson fluctuations. We repeat the full line-search and upper-limit procedure for 1000 mock realizations, following a strategy similar to Refs.~\cite{Fermi-LAT:2015kyq, Abdo:2010nc}. From the resulting distribution of upper limits at each energy, we derive the median expected limit together with the 68\% and 95\% containment bands. Figure~\ref{fig:limitswithbands} shows the observed limit together with these expected sensitivity bands.

In Fig. \ref{fig:boundscomparison} we compare the limits from our subhalo analysis with previous gamma-ray line constraints from the Galactic Center obtained by the Fermi-LAT Collaboration~\cite{Fermi-LAT:2015kyq}, Foster et al.~\cite{Foster_2023} and De La Torre et al.~\cite{luque2023gammaraylines15years}. Our results are generally stronger over most of the mass range. This comparison should not be interpreted as a direct analysis improvement, since the targets, J-factors and background levels are different. Instead, the stronger limits are mainly driven by the astrophysical normalization of the target: if the candidate is interpreted as a DM subhalo, its proximity and mass imply a larger J-factor than those associated with the regions commonly used in GC line searches. This makes the candidate particularly interesting for indirect DM detection strategies.

\begin{figure}[!t]
    \centering
    \includegraphics[width=1\linewidth]{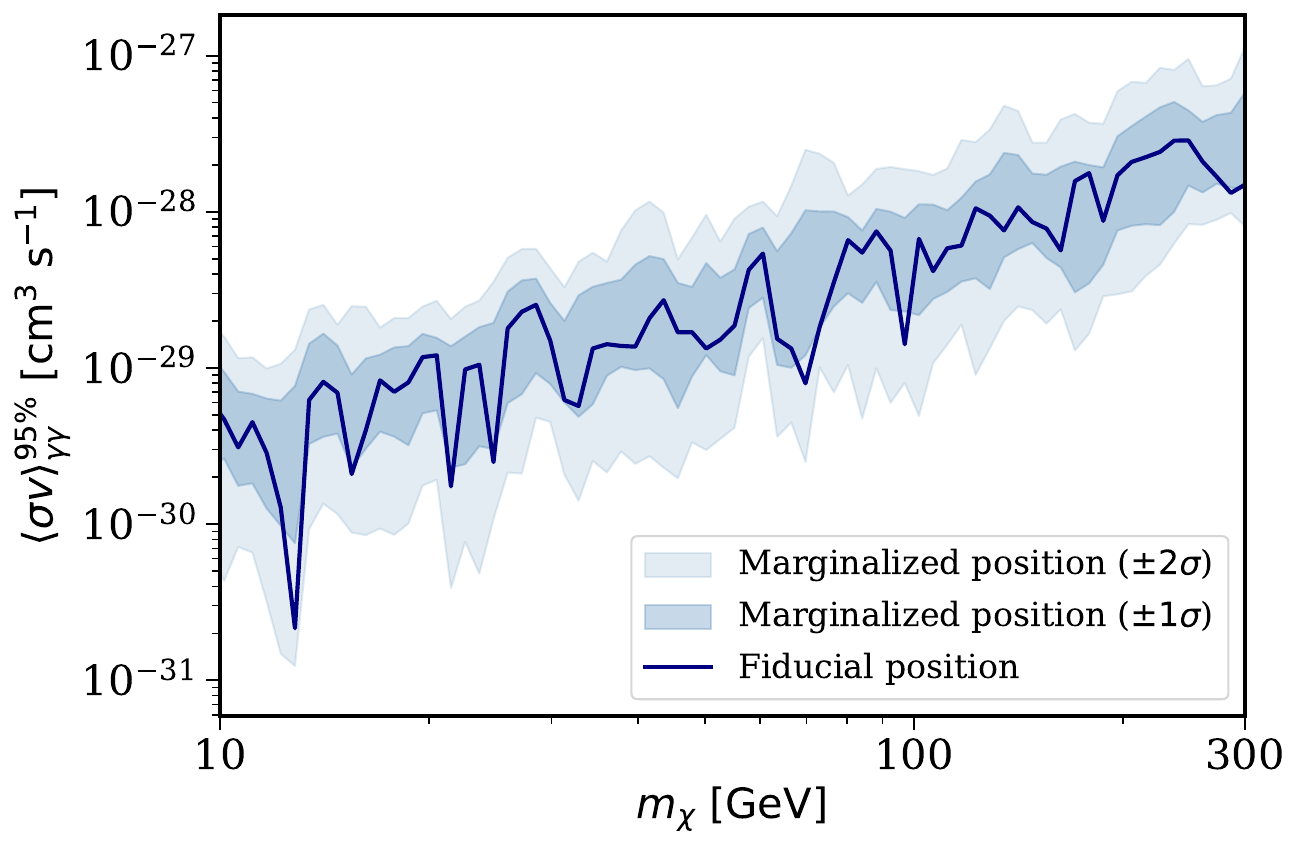}
    \caption{95\% confidence level upper limits on the DM annihilation cross section $\langle\sigma v\rangle_{\gamma\gamma}$. The solid line shows the result for the fiducial subhalo position. The shaded bands represents the positional uncertainty, obtained by marginalizing over 300 random positions within its $2\sigma$ 3D Galactocentric coordinate values. The darker blue encloses the $\pm1\sigma$ while the lighter one the $\pm2\sigma$ of the distribution }
    \label{fig:boundssystem}
\end{figure}

Unlike GC analyses, where the interpretation is affected by a high astrophysical background, unresolved sources, and uncertainties in the Galactic DM density profile, the leading astrophysical uncertainty in our case is associated with the subhalo position. This is shown in Fig. \ref{fig:boundssystem}. As mentioned above, we account for the positional uncertainty using a weighted scan of 300 subhalo positions within the $2\sigma$ region. We repeat the full limit analysis for each and derive the $1\sigma$ and $2\sigma$ bands with the weighted distribution of individual limits.

\subsection{Implications on well-motivated Dark Matter models}
\label{ssec:DMmodels}

The limits derived above apply to $\langle \sigma v \rangle_{\gamma\gamma}$ directly and make no assumption regarding the underlying DM model. Realistic WIMP candidates, however, often do not annihilate into photons at tree level. The $\gamma\gamma$ final state usually arises at one loop, and is therefore strongly suppressed relative to the channels that carry the bulk of the annihilation rate. The size of this suppression depends on the model, but within any given model it introduces no new free parameters: the ratio of the line rate to the total annihilation rate within a given model follows from the particle content and measured SM quantities alone. We can therefore convert our line constraints of Fig.~\ref{fig:boundscomparison} into model-dependent bounds on the total annihilation cross section $\langle \sigma v \rangle_{\rm ann}$ which sets the thermal relic abundance and which continuum searches constrain directly. Because a monochromatic line sits on a nearly featureless astrophysical background, a strong limit on a small branching ratio still competes with searches for the far larger continuum signal. We apply this reasoning to several well-motivated DM models, focusing on the Higgs portal and on the nearly pure wino limit of supersymmetric neutralino DM, both of which have been proposed as explanations of the Galactic Center Excess (GCE). 

In both cases, we compare our constraints with previous constraints on any putative $\gamma$-ray line at the Galactic Center, as in Ref.~\cite{Foster_2023}. Notably, both Higgs Portal and Neutralino models with dark matter masses in the range of approximately 40--100~GeV are motivated by observations of the Galactic Center Excess, which may be produced by the continuum emission of these classes of dark matter models. For this reason, Galactic line constraints for Higgs Portal and Neutralino models are often quoted using an adiabatically contracted dark matter profile with $\gamma \approx$ 1.25, a value that is motivated by Galactic Center Excess observations~\cite{Daylan:2014rsa}. Our subhalo constraints, would not be subject to adiabatic contraction, due to the lack of any central baryonic core. Thus, we conservatively compare our subhalo constraints on dark matter lines using a standard NFW profile against the more optimistic Galactic Center profile choice of $\gamma \approx$ 1.25, which is shown in Refs.~\cite{luque2023gammaraylines15years, Foster_2023}.

\subsubsection{Higgs Portal Dark Matter}
\label{sssec:HiggsPortal}

The Higgs Portal is the most economical way to connect a dark sector to the SM~\cite{Arcadi:2019lka}. A singlet DM field carries no gauge charge and therefore cannot couple to SM matter through the gauge interactions, but it can couple to the Higgs at the renormalizable level, which promotes the Higgs boson to the sole mediator between the two sectors~\cite{Patt:2006fw}. This minimality buys a great deal of predictivity. SM Higgs physics fixes the branching ratios into the annihilation final states, and the requirement of reproducing the observed relic abundance fixes the DM-Higgs coupling, so the DM mass remains the only free parameter~\cite{Chu:2011be}. The scenario has been studied extensively as an explanation of the GCE: for $m_\chi \lesssim 50$~GeV a Higgs-mediated annihilation proceeds dominantly into $b\bar{b}$, and the resulting continuum spectrum reproduces the observed excess well. The same coupling that sets the annihilation rate also contributes to the invisible Higgs width, so collider measurements constrain the model directly, and the Higgs Portal becomes one of the few DM candidates that indirect detection and precision Higgs physics can attack simultaneously.

The portal admits two minimal realizations, distinguished by the spin of the DM field.
\begin{itemize}
  \item A real singlet scalar $S$ of mass $m_S$, which after electroweak
  symmetry breaking couples to the Higgs as
  \begin{equation}
    \mathcal{L} \supset -\frac{1}{2} m_S^2 S^2
      - \frac{\lambda_p v_H}{2}\, h S^2 \, ,
    \label{eq:hp_scalar}
  \end{equation}
  where $\lambda_p$ is a dimensionless coupling and $v_H$ is the Higgs vacuum
  expectation value~\cite{Silveira:1985rk, McDonald:1993ex, Burgess:2000yq}. The spin-independent scattering cross section is unsuppressed here, and current direct detection limits admit only
  $\lambda_p \lesssim 10^{-3}$~\cite{Duerr:2015aka, LZ:2022lsv}, which confines a GCE explanation to a narrow mass window around the Higgs resonance~\cite{Duerr:2015bea}.
  
  \item A Majorana fermion $\chi$ of mass $m_\chi$, coupled to the Higgs
  through a pseudoscalar operator,
  \begin{equation}
    \mathcal{L} \supset -\frac{1}{2} m_\chi \bar{\chi}\chi
      - i\,\frac{y_p}{2}\, h\, \bar{\chi}\gamma_5 \chi \, ,
    \label{eq:hp_majorana}
  \end{equation}
  where $y_p$ is dimensionless. The pseudoscalar structure suppresses the
  spin-independent scattering cross section for real $y_p$~\cite{Fraser:2020dpy}, so
  direct detection experiments are blind to this realization and only
  colliders and indirect searches can probe it.
\end{itemize}

Both realizations remain subject to the bound on invisible Higgs decays,
${\rm BR}(h \to {\rm inv.}) < 0.11$~\cite{ATLAS:2022yvh}, which is difficult to evade. For the case of subhalo observations with a photon energy range from 10--300~GeV, we focus on the Majorana DM scenario with a wider mass range for the GCE fitting. We note that our gamma-ray line constraints can also be converted into total cross-section constraints for the scalar DM scenario.

\begin{figure}[!t]
    \centering
    \includegraphics[width=1\linewidth]{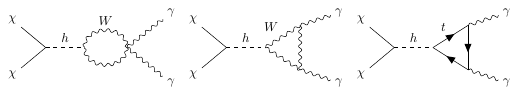}
    \caption{Loop-induced diagrams of dominant annihilation channels to $\gamma$-ray line signal for Higgs Portal model with Majorana DM.}
    \label{fig:HiggsDiagrams}
\end{figure}

Because the annihilation proceeds through an off-shell Higgs, the final state composition is simply that of a SM Higgs of mass $2m_\chi$. The continuum flux therefore follows the Higgs branching ratios evaluated at $m_h = 2m_\chi$, which for the masses of interest are dominated by $b\bar{b}$ with a subdominant but spectrally harder $\tau^{+}\tau^{-}$ component. The $\gamma\gamma$ final state arises only at one loop, through the $W$ and top diagrams of Fig.~\ref{fig:HiggsDiagrams}, and the same argument fixes its rate relative to the total,
\begin{equation}
  \frac{\langle \sigma_{\chi\chi\to\gamma\gamma} v \rangle}
       {\langle \sigma v \rangle_{\rm ann}}
  = \left. {\rm BR}_{h\to\gamma\gamma}\right|_{m_h = 2m_\chi} \, ,
  \label{eq:hp_ratio}
\end{equation}
which is of order $10^{-3}$ for $m_\chi \sim 50$~GeV~\cite{LHCHiggsCrossSectionWorkingGroup:2016ypw}. We take the
branching ratios from Refs.~\cite{Foster_2023, luque2023gammaraylines15years}. No free parameter
enters Eq.~\eqref{eq:hp_ratio}, since the portal coupling cancels between numerator and denominator, so our line limits convert into bounds on the total annihilation rate with no further model input.

\begin{figure}[!t]
    \centering
    \includegraphics[width=1\columnwidth]{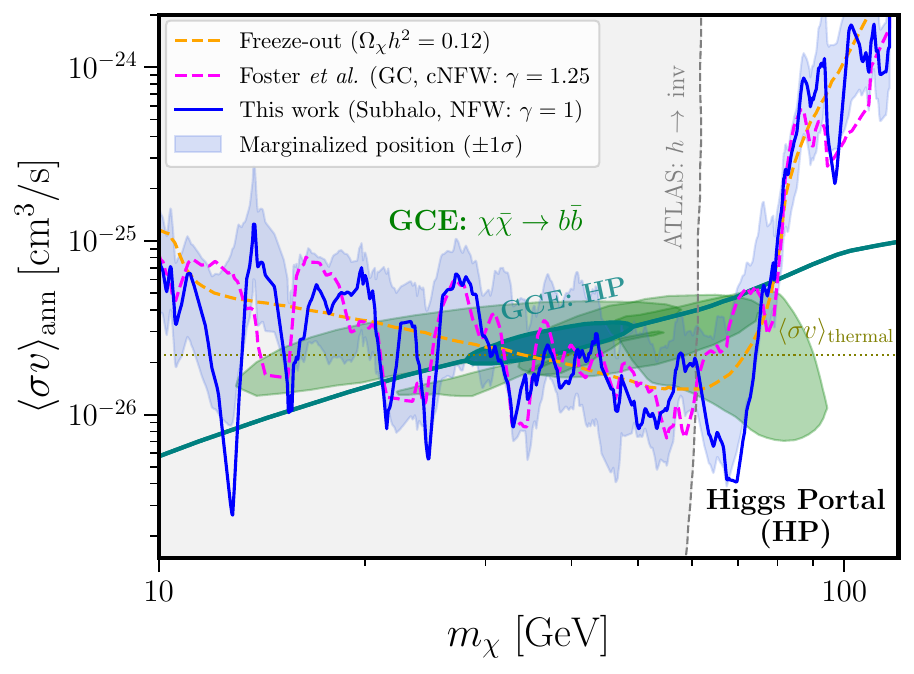}
    \caption{Constraints on the total DM annihilation cross section in the Higgs Portal DM model. Our subhalo constraint, assuming an NFW profile, is the solid blue line, with a $1\sigma$ uncertainty band that stems from marginalizing over subhalo position. The corresponding Galactic Center constraint on the same gamma-ray line signal is shown for an adiabatically contracted NFW profile ($\gamma=1.25$), as the dashed magenta line~\cite{Foster_2023}. The ATLAS invisible Higgs decay bound is in gray~\cite{ATLAS:2022yvh}, and the freeze-out benchmark for $\Omega_\chi h^{2} = 0.12$ is the dashed orange line, with the thermal value \mbox{$\langle \sigma v\rangle_{\rm ann} = 2.2\times 10^{-26}$~cm$^{3}$/s} in dotted olive. For the GCE, the combined $\chi\chi \to b\bar{b}$ fits are in green shades~\cite{Goodenough:2009gk, Hooper:2011ti, Abazajian:2012pn, Gordon:2013vta, Daylan:2014rsa, Calore:2014xka, Calore:2014nla} and are shown for the best fit value of $\gamma$ found in each individual result. The Higgs Portal fit is in teal: the line is the best fit at each fixed $m_\chi$~\cite{Foster_2023, DiMauro:2021qcf}, while the nested ellipses are the $68\%$ and $95\%$ CL regions with both $m_\chi$ and $\langle \sigma v \rangle_{\rm ann}$ free~\cite{Foster_2023, luque2023gammaraylines15years}, which is shown at a self-consistent adiabatically contracted NFW profile of $\gamma=1.25$.}
    \label{fig:higgs}
\end{figure}

Fig.~\ref{fig:higgs} shows the subhalo constraints on the total thermally averaged DM annihilation cross section in the Higgs Portal DM model, assuming the canonical NFW profile. The solid blue line is our subhalo limit, and the band around it covers the $1\sigma$ spread we obtain by marginalizing over the subhalo position. The dashed magenta line gives previous limits from the Fermi-LAT line-search analysis of Galactic Center observations in Ref.~\cite{Foster_2023}\footnote{Ref.~\cite{luque2023gammaraylines15years} also obtained limits on the Higgs Portal model from Galactic Center line searches, of similar magnitude. We compare our results with Ref.~\cite{Foster_2023} only to highlight the difference between Galactic Center and subhalo searches, and our conclusions apply equally to Ref.~\cite{luque2023gammaraylines15years}.}. For Galactic Center observations, we assume a steeper adiabatically contracted NFW profile with  $\gamma=1.25$, which is motivated by the prospect that such DM models can fit the morphology of the observed Galactic Center Excess.  The dashed orange line traces the cross section that reproduces the correct relic abundance $\Omega_{\chi} h^{2} = 0.12$ through thermal freeze-out, and the dotted olive line marks the canonical thermal value
\mbox{$\langle \sigma v \rangle_{\rm thermal} = 2.2\times 10^{-26}$~cm$^{3}$/s}. ATLAS searches for invisible Higgs decays exclude the gray region, which terminates at $m_{\chi} = m_{h}/2$ (gray dashed line)~\cite{ATLAS:2022yvh}. The green shades cover the GCE parameter space from fits that assume pure $\chi\chi \to b\bar{b}$ annihilation~\cite{Goodenough:2009gk, Hooper:2011ti, Abazajian:2012pn, Gordon:2013vta, Daylan:2014rsa, Calore:2014xka, Calore:2014nla}, while the teal line and ellipses specialize to the Higgs Portal: the line follows the best-fit cross section at each fixed $m_{\chi}$~\cite{Foster_2023, DiMauro:2021qcf}, and the ellipses enclose the 68\% and 95\%~CL regions with both $m_{\chi}$ and $\langle \sigma v \rangle_{\rm ann}$ free~\cite{Foster_2023, luque2023gammaraylines15years}.

We can divide our constraints into two mass regimes. Below $m_{h}/2$, the ATLAS constraints on invisible Higgs decay generically apply. In this regime, our subhalo limits are comparable to those from  previous Galactic Center line searches~\cite{Foster_2023, luque2023gammaraylines15years}. Our constraints dip below the canonical thermal value \mbox{$\langle \sigma v \rangle_{\rm thermal}$} and the thermal freeze-out benchmark over part of this mass range. They also cut into the region preferred by GCE fits to the $b\bar{b}$ annihilation channel~\cite{Goodenough:2009gk, Hooper:2011ti, Abazajian:2012pn, Gordon:2013vta, Daylan:2014rsa, Calore:2014xka, Calore:2014nla} and into the best-fit region in which the Higgs Portal model explains the GCE~\cite{Foster_2023, DiMauro:2021qcf}. Notably, we reach this sensitivity assuming only a standard NFW profile, which demonstrates that a nearby subhalo can probe these benchmarks without invoking a contracted inner-Galaxy profile, and our $J$-factor follows from the dynamically inferred mass and distance of the source rather than from an assumed density profile. While ATLAS searches for invisible Higgs decays already cover this parameter space and typically rule out thermal dark matter models~\cite{ATLAS:2022yvh}, those constraints can be relaxed once extra degrees of freedom enter, as in non-minimal realizations of the portal~\cite{Mondal:2014goi, Bell:2017rgi, Ipek:2014gua, Yang:2018fje, Cuoco:2016jqt}. Our limits continue to apply even where the Higgs decay bounds do not.

For masses above $m_{h}/2$, the invisible decay channel closes and the ATLAS bound disappears entirely. Since the pseudoscalar coupling also suppresses direct detection through a spin-dependent interaction, indirect detection supplies the only constraint on the Higgs Portal in this regime. Thus, in this mass range the result is not merely complementary to collider probes but the sole probe available. The freeze-out benchmark rises steeply just above the resonance, because the late-time thermally averaged cross section departs from its early-universe value near $m_{h}/2$. A thermal Higgs Portal candidate above $m_{h}/2$ therefore annihilates efficiently today, which makes it a natural target for line searches, and our limits approach that benchmark between 60--85~GeV. Our subhalo limits also push the exclusion of the freeze-out benchmark to higher masses than the Galactic Center line search reaches, and they remove the part of the $b\bar{b}$ GCE region that extends above $m_{h}/2$~\cite{Calore:2014nla, Calore:2014xka}. The constraint weakens rapidly above $m_{\chi} \simeq 85$~GeV, where $WW$ and $ZZ$ final states open for an off-shell Higgs of mass $2m_{\chi}$ and ${\rm BR}_{h\to\gamma\gamma}$ drops, suppressing the line relative to the total annihilation rate.

\subsubsection{Wino Dark Matter}
\label{sssec:winoDM}

The Wino is the $SU(2)$ triplet fermion of supersymmetric extensions of the SM, the superpartner of the $W$ boson~\cite{Martin:1997ns}. The lightest Neutralino is, in general, a mixture of the Bino, the Wino, and the two Higgsinos, and we call it Wino-like when the Wino component dominates the mixture~\cite{Ellis:1983ew, Jungman:1995df, Roszkowski:2017nbc}.  The Neutralino $\chi$ acquires a nearly degenerate charged partner, the Chargino $\chi^{\pm}$. Therefore, the Neutralino annihilates dominantly into $W^{+}W^{-}$ through $t$-channel Chargino exchange. Neutralino DM draws independent motivation from the electroweak hierarchy problem, which favors supersymmetry near the electroweak scale~\cite{Wells:2003tf}. The Wino also matches the predictivity of the Higgs Portal model, for a different reason: it couples to the SM through gauge interactions alone, so no free coupling survives and the annihilation rate follows entirely from Neutralino and Chargino masses. Following Ref.~\cite{Foster_2023}, we adopt a Split-SUSY spectrum~\cite{Giudice:2004tc, Arkani-Hamed:2004ymt, Arvanitaki:2012ps, Arkani-Hamed:2012fhg}, which decouples the sfermions and leaves the electroweak gauge bosons as the dominant annihilation channel. We model the dark sector as an approximately pure Wino, a Majorana Neutralino $\chi$ together with its nearly degenerate Chargino.

The lightest Neutralino provides a viable DM candidate~\cite{Martin:1997ns, Arkani-Hamed:2006wnf},
with a phenomenology that follows from the gauge couplings alone,
\begin{equation}
    \mathcal{L} \supset -g\, W^{\mp}_{\mu}\, \bar{\chi}\gamma^{\mu}\chi^{\pm}
  \;-\; e\, A_{\mu}\, \bar{\chi}^{\pm}\gamma^{\mu}\chi^{\pm} \, ,
  \label{eq:wino_lagrangian}
\end{equation}
where $g$ and $e$ are the weak and electromagnetic couplings. At tree level the Wino annihilates to $W^{+}W^{-}$, and for $m_{\chi}\sim 100$~GeV the rate evaluates to \mbox{$\langle\sigma v \rangle \approx 4\times 10^{-24}$~cm$^{3}$/s}~\cite{Foster_2023}, two orders of magnitude above the thermal value. Such efficient annihilation prevents the Wino from freezing out as all of the DM, so a standard thermal history leaves only a small fraction, $f_{\chi}\sim 10^{-2}$, and suppresses the annihilation signal by $f_{\chi}^{2}$. Non-thermal production instead allows any fraction up to the whole, so we consider both scenarios.

\begin{figure}[t]
    \centering
    \includegraphics[width=1\columnwidth]{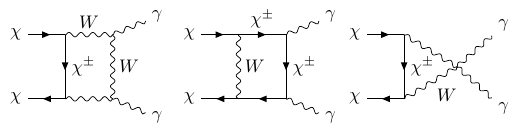}
    \caption{Loop-induced diagrams of dominant annihilation channels to $\gamma$-ray line signal for Wino DM model.}
    \label{fig:winoDiagrams}
\end{figure}

Many studies have considered Neutralino DM in the context of indirect detection~\cite{Agrawal:2014oha, Achterberg:2015srl, Murgia:2020dzu, Butter:2016tjc}. For the Wino DM scenario, annihilation into $\gamma\gamma$ arises only at one loop, through the $W$ and Chargino diagrams in Fig.~\ref{fig:winoDiagrams}. The annihilation cross section reaches \mbox{$\langle \sigma v \rangle_{\gamma\gamma}\sim 10^{-2}\langle \sigma v \rangle_{WW}$} near 100~GeV~\cite{Foster_2023}. The Chargino therefore controls both the continuum and the line, and its mass splitting from the Neutralino requires care. Radiative corrections set that splitting to
\mbox{$\Delta m_{+}\sim150$--$160$~MeV} in the pure Wino limit~\cite{Ibe:2012sx}. However, collider searches exclude Charginos below 270~GeV with \mbox{$\Delta m_{+}\lesssim 220$~MeV}~\cite{ATLAS:2013ikw}, and LEP disfavors \mbox{$m_{\chi^{\pm}}\lesssim 95$~GeV} outright~\cite{DELPHI:2003uqw, ALEPH:2002gap}. Following Ref.~\cite{Foster_2023}, we therefore relax the splitting to \mbox{$\Delta m_{+}=0.2m_{\chi}$}, which keeps \mbox{$m_{\chi^{\pm}}>95$~GeV} across the mass range we consider. We attribute this larger splitting to a small Bino admixture that leaves the annihilation phenomenology intact. We truncate the analysis at $m_{\chi}=m_{W}$, below which the $WW$ channel closes. The fraction of the DM that the Wino constitutes does not enter the ratio between the continuum and the line, since the same DM produces both, so a line limit constrains the observed annihilation signal regardless of which scenario holds.

\begin{figure}[!t]
    \centering
    \includegraphics[width=1\columnwidth]{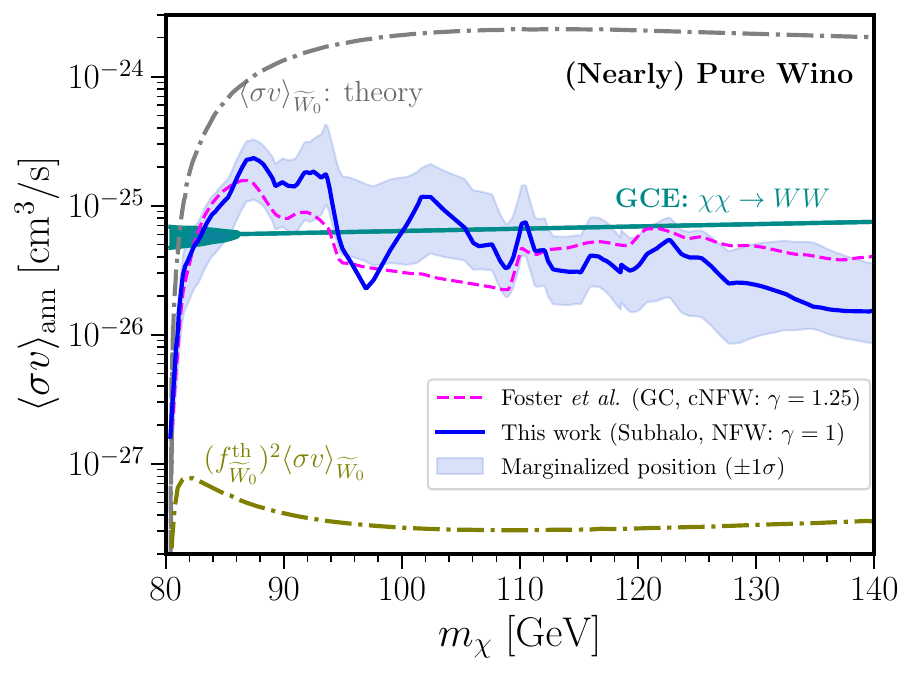}
    \caption{Constraints on the total thermally averaged annihilation cross section of Wino DM. Our subhalo constraint using an NFW profile is the solid blue line, with its $1\sigma$ marginalized position band. Previous Galactic Center constraint with an adiabatic contracted NFW profile ($\gamma=1.25$) is the dashed magenta line~\cite{Foster_2023}. The solid teal line is the best-fit cross section to the GCE at each fixed DM mass, and the teal ellipses are the 68\% and 95\% containment regions with both $m_{\chi}$ and the cross section free~\cite{Foster_2023}. Two theoretical benchmarks appear as dot-dashed lines, both for a Neutralino-Chargino mass gap of \mbox{$\Delta m_{+} = 0.2 m_{\chi}$}: first, a scenario where the Wino makes up all of the DM (gray), and second, a scenario where the Wino carries only its thermal fraction, with the cross section multiplied by the sub-fraction squared (olive).}
    \label{fig:wino}
\end{figure}

Fig.~\ref{fig:wino} shows constraints on the total annihilation cross section of Wino DM. The solid blue line gives our subhalo limit, and the band around it covers the $1\sigma$ spread we obtain by marginalizing over the subhalo position. The dashed magenta line gives the corresponding limit from the Galactic Center line search with an adiabatic contracted NFW profile ($\gamma=1.25$)~\cite{Foster_2023}. The solid teal line traces the best-fit cross section to the GCE at each fixed $m_{\chi}$, and the teal ellipses enclose the 68\% and 95\% CL regions with both $m_{\chi}$ and the cross section free. These contours appear as half ellipses because the GCE fit prefers a near-threshold wino and we truncate at $m_{\chi}=m_{W}$~\cite{Agrawal:2014oha, Achterberg:2015srl}. Two dot-dashed lines give the theoretical wino cross section for \mbox{$\Delta m_{+}=0.2m_{\chi}$}: the gray line assumes the Wino makes up all of the DM, and the olive line scales that prediction by the squared thermal sub-fraction $f_{\chi}^{2}$.

Our subhalo limits are comparable to the Galactic Center search over most of the mass range, and improve on it above $\sim$110~GeV. The theoretical cross section for a Wino that makes up all of the DM is excluded over our full mass range, so a Wino between \mbox{80--140~GeV} cannot constitute the whole DM abundance. The GCE best-fit cross section also lies above our limit except in a narrow window between \mbox{84--92~GeV}, where our limit weakens. The parameter space the GCE most prefers, delimited by the containment ellipses, lies above our bound between \mbox{80--84~GeV} but falls below it between \mbox{84--86~GeV}, so we exclude roughly two thirds of that region while the remainder survives once we include the $1\sigma$ positional uncertainty. A thermally produced Wino instead sits orders of magnitude below our sensitivity once the $f_{\chi}^{2}$ suppression applies, and line searches leave it untouched. Our results nevertheless show that gamma-ray lines from nearby subhalos can constrain a Wino origin for the GCE over most of the mass range where such an explanation has been proposed\footnote{Ref.~\cite{Foster_2023} also applies this line search to the Galactic Center to investigate the Higgsino case. However, a thermal Higgsino that reproduces the observed relic abundance requires $m_{\chi}\sim1.1$~TeV, while our subhalo line search extends only to 300~GeV, so we do not consider it here.}, and do so assuming only a standard NFW profile and without any assumption about the Wino production history.

\section{CONCLUSIONS}
\label{sec:conclusions}

In this work, we have performed a dedicated search for monochromatic gamma-ray lines within a nearby DM subhalo candidate. Using nearly 15 years of Fermi-LAT Pass 8 data, we looked for line-like features between 10--300 GeV with a sliding window likelihood analysis. We find no statistically significant evidence for a gamma-ray line. The largest excess appears around $E_\gamma \simeq 28 \, {\rm GeV}$ and reaches a local significance of approximately $2.5\sigma$. The corresponding off-regions do not show a comparable feature at the same energy, and the excess remains visible over our marginalized distribution of potential subhalo locations. Nevertheless, its modest value after trial factors are accounted for does not constitute statistically significant evidence for any dark matter signal. 

We therefore use the absence of a significant line detection to place 95\% confidence level upper limits on $\langle \sigma v\rangle_{\gamma\gamma}$. Background-only pseudo-experiments provide the expected sensitivity of the analysis and show that the observed limits remain broadly compatible with statistical fluctuations. A central aspect of this target is the uncertainty in its inferred position and consequently in its astrophysical J-factor. We account for this uncertainty by doing a weighted scan of 300 realizations of the Galactocentric position within the $2\sigma$ values of each position coordinate. For each location, we repeat the full upper limit analysis and use the weighted distribution to define the $1\sigma$ and $2\sigma$ positional-uncertainty bands around the limit coming from the fiducial position. 

We note that the uncertainty in the subhalo location produces a non-negligible spread in the annihilation constraints, but does not alter the main conclusions of the analysis. Notably, most of this uncertainty stems from the distance uncertainty of the object, and its effect on the measured $J$-factor. Our analysis is highly robust to changes in the angular position and extent of the object, even over angular scales that reach tens of degrees. This demonstrates the power of using $\gamma$-ray line searches, compared to continuum searches, in scenarios where the location and profile of a dark matter structure are highly uncertain. In any continuum search, the large number of unassociated $\gamma$-ray sources in a region of interest will always impart a significant systematic uncertainty which makes it difficult to claim any robust constraint~\cite{zhu2025constraintsdarkmatterannihilation}. The spectral specificity of the line search avoids these systematic issues, and limits the astrophysical background to a primarily statistical uncertainty that is less sensitive to changes in the objects position or extent.

Within the context of models where both the Galactic Center and subhalo dark matter density profiles are fit by similar NFW profiles, our limits on $\langle \sigma v\rangle_{\gamma\gamma}$ are stronger than previous constraints from Galactic Center line searches over most of the explored energy range. This comparison should not be regarded as a direct improvement of these previous works, since we are exploring a different, and more uncertain, target. Instead, it highlights the potential of dynamically selected nearby DM substructure for indirect searches, whose proximity can lead to large $J$-factors.

We also translate our subhalo model-independent limits into constraints on well-motivated particle DM scenarios. For Higgs Portal DM, the subhalo limits are comparable to previous Galactic Center constraints on the total annihilation cross section, and improve on them in a window extending from $m_h/2$ to roughly 73~GeV. They reach the thermal freeze-out benchmark over part of the range below the Higgs resonance and cut into the region in which Higgs Portal annihilation could account for the Galactic Center Excess. They remain particularly valuable above $m_h/2$, where the invisible Higgs-decay constraint no longer applies. For Wino DM, our constraints exclude a Wino that constitutes the full DM abundance over the \mbox{80--140~GeV} range and constrain most of the parameter space in which it could explain the Galactic Center Excess, while a thermally produced Wino carrying only its expected sub-fraction of the DM remains below the sensitivity of this search. We reach this sensitivity while conservatively modeling the subhalo with a standard NFW profile, whereas Galactic Center analyses adopt a modestly contracted profile for the inner Galaxy, so our constraints provide complementary constraints that are independent of the assumed dark matter density profile in the Galactic Center.

The main astrophysical uncertainty comes from the subhalo position and the corresponding $J$-factor. We note that an additional uncertainty arises from the poorly known internal density profile of the subhalo. We adopt the standard NFW profile in order to be consistent with Ref.~\cite{29xz-nt5z} and as a conservative benchmark. Notably, tidal disruption can substantially reshape the density profile of these objects, usually removing material from their outer regions while leaving cuspy central mass remnant, similar to a truncated NFW profile that maintains its nucleus \cite{Errani:2020wgn, Green:2019zkz}, which could modify the $J$-factor and consequently changing our annihilation constraints. Improved pulsar-timing measurements and additional Fermi-LAT exposure would therefore strongly improve both the astrophysical normalization and the statistical sensitivity. Additionally, the dark matter constraints at higher masses may be improved through novel searches for the double-line feature produced by the coupled annihilation to $\gamma\gamma$ and Z$\gamma$ final states~\cite{luque2023gammaraylines15years} when relying on specific particle models. Finally, the same target also provides a natural opportunity to test other exotic DM scenarios~\cite{Fuller:2024noz, Ibarra:2007wg, Carenza:2023eua, Dienes:2014via, Nguyen:2024kwy, Nguyen:2025tkl, Garny:2010eg}, including decaying dark matter through its $D$-factor. We leave these for future work.

Overall, our results demonstrate the power of using gamma-ray line searches to probe dark matter annihilation in nearby DM substructure that has been gravitationally selected. If future dynamical observations strengthen the subhalo interpretation of this candidate, its proximity and large inferred J-factor make it an exceptionally sensitive target for indirect detection. Our ability to further marginalize our subhalo constraints over the uncertain position of these subhalos implies that such sources could produce world-leading constraints on DM particle physics, even in scenarios where we lack any non-gravitational detection of the subhalo itself.

\section*{Acknowledgements}
  ASP acknowledges support from the European traineeship programme after master studies funded through Erasmus+ in Universidad Autónoma de Madrid (UAM), which enabled a research stay in Stockholm, and thanks Tim Linden and his group for their hospitality during this visit. TTQN thanks the TASI organizers and the University of Colorado Boulder, as well as the Instituto de Astrof\'{\i}sica de Canarias (IAC) for their hospitality. TTQN and TL are supported by the Swedish Research Council under contract 2022-04283. TTQN is also supported by two grants from the Royal Swedish Academy of Sciences (KVA): PH2025-0073 (Physics) and AST2025-0048 (Astronomy and Space Science).
  PDL has been supported by the Juan de la Cierva JDC2022-048916-I grant, funded by MCIU/AEI/10.13039/501100011033 European Union "NextGenerationEU"/PRTR, and is currently supported by Ramón y Cajal RYC2024-048445-I grant, which is funded by MCIU/AEI/10.13039/501100011033 and FSE+. The work of PDL is also supported by the grants PID2021-125331NB-I00 and CEX2020-001007-S, both funded by MCIN/AEI/10.13039/501100011033 and by ``ERDF A way of making Europe''. PDL also acknowledges the MultiDark Network, ref. RED2022-134411-T. Most of the calculations performed in this work were done using computing resources provided by the National Academic Infrastructure for Supercomputing in Sweden (NAISS) under the project 2025/5-729, which is partially funded by the Swedish Research Council through Grant 2022-06725.

\appendix

\section{Additional plots}
\label{sec:additionalp}
\begin{figure}[tbp]
    \centering
    \includegraphics[width=1\linewidth]{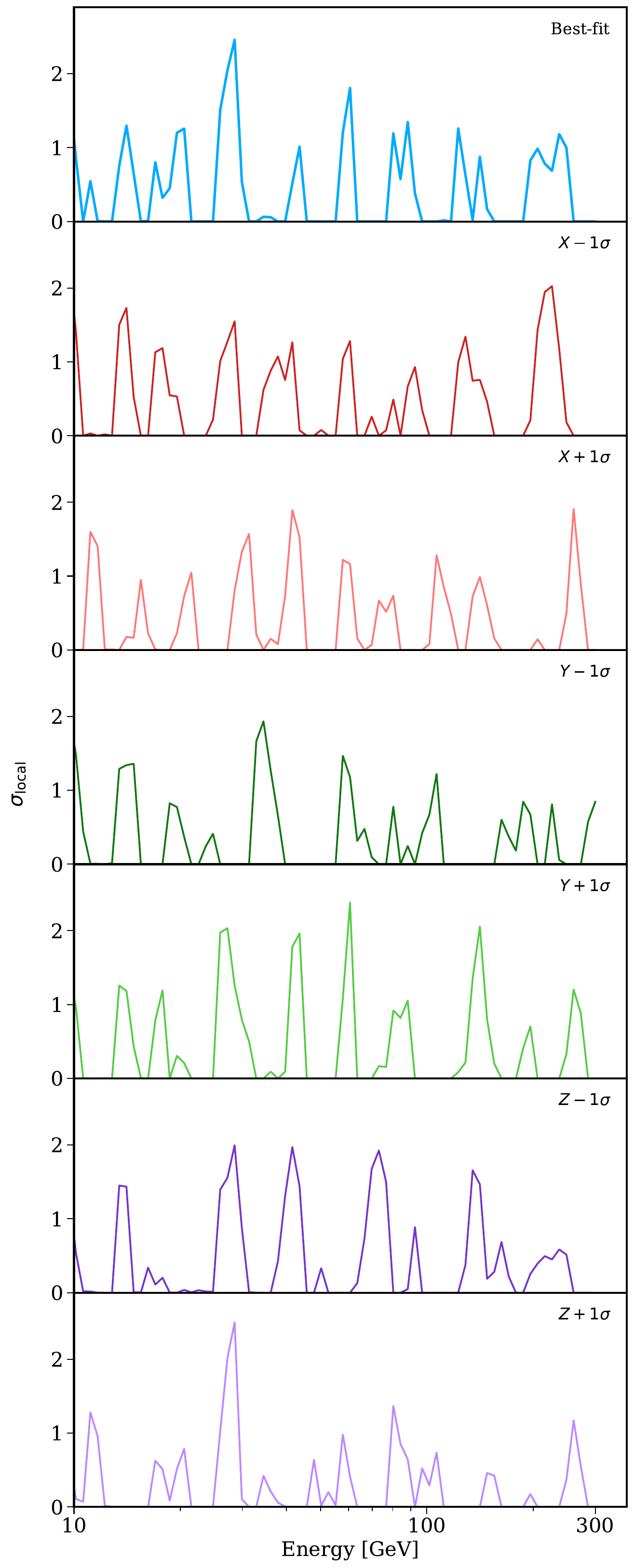}
    \caption{Local significance obtained in the 10--300 GeV range for the fiducial subhalo position and for positions shifted by $\pm1\sigma$ in the Galactocentric coordinates. Each coordinate is varied independently while the remaining coordinates are kept fixed at their fiducial values. The $\sigma_{\rm local}$ is set to 0 when the fitted line normalization is not compatible with a positive number.}
    \label{sigmalocations}
\end{figure}

As an additional robustness check, we repeat the line search after shifting the subhalo position within the quoted $1\sigma$ uncertainties of its Galactocentric coordinates. Each coordinate is varied independently by $\pm1\sigma$ while keeping the remaining coordinates fixed at their fiducial values. The resulting local significances are shown in Fig.~\ref{sigmalocations}.

The feature near $E_\gamma\simeq28$ GeV remains visible in several of the shifted configurations, in some cases reaching local significances above $2\sigma$, although its strength varies with the assumed position. This indicates that the feature is not restricted to the exact fiducial coordinates of the subhalo. Its significance nevertheless remains statistically modest in all configurations. These shifted positions are used only as a check; the positional uncertainty on the annihilation limits is instead evaluated using the 300-position weighted scan described in the main text.

\clearpage
\bibliography{main}

\end{document}